# Supercharging Recommender Systems using Taxonomies for Learning User Purchase Behavior

Bhargav Kanagal* Amr Ahmed * Sandeep Pandey Vanja Josifovski Jeff Yuan Lluis Garcia-Pueyo

{kbhargav*, amrahmed*, spandey, vanjaj, yuanjef, lluis}@yahoo-inc.com

Yahoo! Research, USA

*co-first authors

## ABSTRACT

Recommender systems based on latent factor models have been effectively used for understanding user interests and predicting future actions. Such models work by projecting the users and items into a smaller dimensional space, thereby clustering similar users and items together and subsequently compute similarity between unknown user-item pairs. When user-item interactions are sparse (*sparsity* problem) or when new items continuously appear (*cold start* problem), these models perform poorly. In this paper, we exploit the combination of taxonomies and latent factor models to mitigate these issues and improve recommendation accuracy. We observe that taxonomies provide structure similar to that of a latent factor model: namely, it imposes human-labeled categories (clusters) over items. This leads to our proposed *taxonomy-aware* latent factor model (TF) which combines taxonomies and latent factors using additive models. We develop efficient algorithms to train the TF models, which scales to large number of users/items and develop scalable inference/recommendation algorithms by exploiting the structure of the taxonomy. In addition, we extend the TF model to account for the temporal dynamics of user interests using high-order *Markov chains*. To deal with large-scale data, we develop a parallel multi-core implementation of our TF model. We empirically evaluate the TF model for the task of predicting user purchases using a real-world shopping dataset spanning more than a million users and products. Our experiments demonstrate the benefits of using our TF models over existing approaches, in terms of both prediction accuracy and running time.

## 1. INTRODUCTION

Personalized recommendation systems are ubiquitous with applications to computational advertising, content suggestions, search engines and e-commerce. These systems use the past behavior of users to recommend new items that are likely to be of interest to them. Significant advancements have been made in recent years to improve the accuracy of such personalized recommendation systems and to scale the algorithms to large amount of data. One of the most extensively used techniques in these recommendation systems is the latent factor model and its variants [4, 17, 18, 20] (see Koren et al. [19] for a survey). The idea behind latent factor models is to project the users and items into a smaller dimensional space (such lower dimensional projections are called *factors*), thereby clustering similar users and items. Subsequently, the interest (similarity) of a user to an unrated item is computed and the most similar item(s) is recommended to the user. While latent factor models have been very successful with Netflix contest top performers [8, 17] and tag recommendation [26] being some of the success stories, these models suffer with certain shortcomings. We illustrate some of these challenges using the example of product recommendation, which is the main focus of the paper. In this application, we need to predict future purchases that will be made by the users, by using the historical purchase logs.

- First, it is difficult to accurately learn the latent factors when the user-item interactions are *sparse*. For instance, in our application, a user has purchased on average, only 2.3 items out of more than a million possible items.
- Second, new items are released continuously and the latent factor model cannot learn the factors for such items, given no data. This is commonly known as the *cold start* problem [4] in recommender systems.
- Third, conventional models do not account for the temporal dynamics in user behavior, e.g., a flash memory is much more likely to be bought soon after a camera purchase. Further, naive temporal-based extensions that condition on previous user purchases introduce additional sparsity (e.g., the number of users who buy flash memory after camera is much fewer than the number of individual purchases for flash memory and camera).

We resolve the aforementioned challenges by using a taxonomy over items which is available for many consumer products, e.g., PriceGrabber shopping taxonomy [3], music taxonomy [16], movie taxonomy (genre, director and so on). A fragment of the Yahoo! shopping taxonomy (that we use in our paper) is shown in Figure 1. Since taxonomies are designed by humans, they capture knowledge about the domain at hand, independent of the training data [13] and hence provide scope for improving prediction. As such, taxonomies can provide lineage for items in terms of categories and their ancestors in the taxonomy, thus they help to ameliorate the *sparsity* issues that are common with online shopping data. Also, taxonomies can help in dealing with the *cold start* problem because even though the set of individual products/items is highly dynamic, the taxonomy is relatively stable. For instance, the ancestors of a newly arrived item can be initially used to guide recommendations for the new item. In addition, taxonomies enable us to provide a *structured ranking*. Traditional latent factor methods provide a ranking over all the (possibly) unrelated products. Using the knowledge of taxonomy enables us to rank items within a given category and rank categories themselves, based on higher levels in the taxonomy. Further, using taxonomies allows us to target users by product categories, which is commonly required in advertising campaigns and reduce duplication of items of similar type.

**Taxonomy-aware dynamic latent factor model.** In this paper, we propose our *taxonomy-aware latent factor model* (TF) which combines taxonomies and latent factors for the task of learning user purchase behavior. We note here that there has been recent literature along this general research direction and provide details comparing our approach with existing approaches in Section 8. In our model, we use the taxonomy categorization as a *prior* and augment this into the latent factor model to provide an *additive* framework for learning. As with latent factor models, we learn factors for users and items to measure the similarity/interest of a user to an item. In addition, we introduce factors for every interior node in the taxonomy and enforce the taxonomy prior over these latent factors: i.e., we learn the factors such that the sibling items in the taxonomy have factors similar to their parent. This also allows items with very few user interactions to benefit from their siblings, thereby leading to better factors. The additive framework also allows items with large number of user interactions to differ from their siblings, if the data exhibits such patterns.

While the above-mentioned factors account for the long-term interests of users in items, they do not explicitly capture the short-term trends that are commonly seen in user online purchase behavior [7]. For instance, if a user bought a camera in one of the last few time steps, then the likelihood that she buys a flash memory or a camera lens is high. We model such short-term trends using *k-order Markov chains* over the sequences in which items are bought one after another. In other words, we learn additional factors for items which capture how likely a given item is bought by a user after purchasing some other item. Note that the taxonomy plays a vital role while building such temporal models since it allows us to condition on the past purchases of the user without introducing additional sparsity. For example, the number of users who bought a "Canon Rebel EOS T1i 15 MP" camera and subsequently purchased a "Sandisk 8GB Extreme SDHC SDSDX3" flash drive is much less than the number of users who bought (any) camera followed by (any) flash drive.

**Model learning and recommendation.** We learn the factors in the model using stochastic gradient descent (SGD), an iterative algorithm for optimization. While the naive algorithm for SGD works for a few hundreds/thousands of items, it does not scale to our shopping dataset (Section 7.1) (requires a large number of iterations). We introduced a novel training algorithm called sibling-based training that enforces user preferences at each level of the taxonomy. Once the model is trained, it is used to make recommendations for the users. Traditional inference methods iterate over all the items for a given user, compute an affinity score for each item to the user, and subsequently compute the ranking over the items based on the score. Since we have over a million items, computing the ranking for all the users in this manner is prohibitively expensive. Instead, we propose a *cascaded* inference algorithm in which we rank in a top-down manner where, for a given user, we first rank the nodes at the highest level of the taxonomy. Then we follow only the $k$ high-scoring nodes and the subtrees under them and continue the recursion. Cascaded inference provides us with a natural accuracy/efficiency trade-off where we can increase the accuracy by increasing $k$ at the expense of efficiency. In addition, by cascading, we can provide a more meaningful structured ranking.

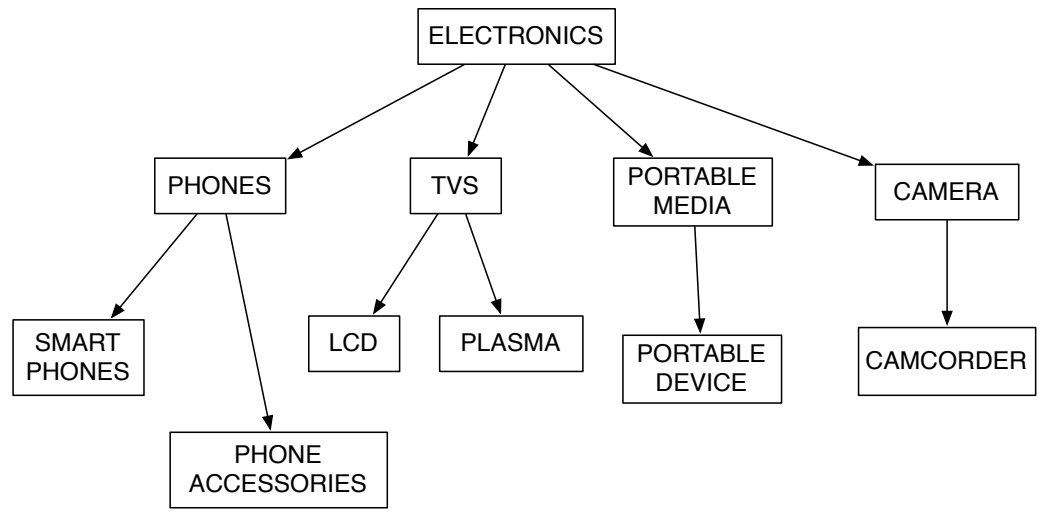

Figure 1: Fragment of the product taxonomy used in Yahoo! Shopping. (Product names omitted)

**Parallel implementation.** We develop a parallel multi-core implementation of the training and inference algorithms. The factor matrices are global and are shared by the various threads. To update a given item factor, a thread needs to obtain a lock on the item factor before writing to it. One of the key challenges we faced with the taxonomy is that item factors are updated with different frequencies by the training algorithm. For instance, the item factors corresponding to the upper levels of the taxonomy are updated more frequently than the items at the leaf of the taxonomy. In order to address this problem, we develop *caching* heuristics that enable us to improve the speedup.

**Contributions.** The contributions of our work include:

1. We propose a novel taxonomy-aware latent factor model (TF), that combines the taxonomies with latent factor models using an additive framework. We show how TF can help alleviate several issues including sparsity and cold start.

2. We introduced a novel algorithm for training taxonomy-guided models that we call sibling-based training and show experimentally that it results in significant improvements.

3. We present efficient algorithms for training TF and executing inference over TF models by exploiting the structure of the taxonomy, called cascaded learning and inferencing.

4. We extend the TF model to differentiate between the long-term and short term purchasing behavior and show it leads to better recommendations.

5. We detail an efficient parallel implementation and empirically demonstrate the benefits of our approach over a real-world large-scale dataset from online shopping sites.

**Outline.** The rest of the paper is organized as follows. We provide background for the various concepts used in the paper in Section 2. We describe our taxonomy-aware latent factor model in Section 3. Next, we discuss techniques for learning TF (Section 4) and making recommendations from TF (Section 5). We provide details of our implementation in Section 6 and conclude with a comprehensive experimental study in Section 7. We briefly mention some of the other related work in Section 8.

# 2. PRELIMINARIES

In this section, we provide background for the various concepts used in the paper. We begin with a brief introduction to latent factor models, a commonly used technique in recommender systems and illustrate *stochastic gradient descent*, a widely used optimization algorithm.

## 2.1 Latent Factor Models

Latent factor models have been extensively used in recommender systems [19, 4]. The most successful techniques in the Netflix prize [8] were based on latent factor models. We illustrate latent factor models below. The input to a recommender system is a sparse (partially populated) user-item matrix $Y$ (size $m \times n$, $m$ users and $n$ items) where the entries correspond to an interaction between a user and an item – either in terms of a rating or a purchase. The goal of the recommender system is to predict, for each user $u$, a ranked list of items missing entries in the matrix. We assume that each user $u$ can be represented by a latent factor $\mathbf{v_u}$ which is a vector of size $1 \times K$ ($K$ is commonly referred to as the number of factors in the model, typically much smaller than $m$ and $n$). Similarly, each item $i$ can be represented by a latent factor $\mathbf{v_i}$ (also, a vector of size $1 \times K$). User $u$'s affinity/interest in item $i$ is assumed to follow this model:

$$\hat{x}_{ui} = \langle \mathbf{v_u}, \mathbf{v_i} \rangle$$

Here, $x_{ui}$ is the affinity of user $u$ to item $i$, $\langle \mathbf{v_u}, \mathbf{v_i} \rangle$ represents the dot product of the corresponding user and item factors. Most latent factor models also include *bias* terms $b_u$ and $b_i$ that model the generosity of the user's rating and the popularity of the item respectively; we ignore this for simplicity of exposition. The learning problem here is to determine the best values for $\mathbf{v_u}$ and $\mathbf{v_i}$ (for all users $u$ and all items $i$) based on the given rating matrix $Y$, these parameters are usually denoted by $\Theta$. While traditional approaches to matrix factorization try to regress over the known entries of the matrix, a more successful approach is the recently proposed Bayesian personalized ranking (BPR) [24]. Here, the trick is to do regression directly over the *ranks* of the items, rather than the actual ratings since the ultimate goal of the system is to construct the ranked lists. Also, we only have *implicit feedback* from the users (i.e., we will have a rating between 1 to 5, but only know that the user made a purchase). In this case, regression over the actual numbers of purchases is not meaningful. In BPR, the goal is to discriminate between items bought by the user and items that were not bought. In other words, we need to learn a ranking function $R_u$ for each user $u$ that ranks u's interesting items higher than the non-interesting items. In other words, if item $i$ appears in user u's purchase list $B_u$ and item $j$ does not appear in $B_u$, then we must have $R_u(i) > R_u(j)$. For this, we need to have: $x_{ui} > x_{uj}$. Based on the above arguments, our likelihood function $p(R_u|\Theta)$ is given by:

$$p(R_u|\Theta) = \prod_{u \in U} \prod_{i \in B_u} \prod_{j \notin B_u} \sigma(x_{ui} - x_{uj})$$

Following Rendle et al. [24], we have approximated the non-smooth, non-differentiable expression $x_{ui} > x_{uj}$ using the logistic sigmoid function $\sigma(x_{ui} - x_{uj})$, where $\sigma(z) = \frac{1}{1+e^{-z}}$. We use a Gaussian prior $N(0, \sigma)$ over all the factors in $\Theta$ and compute the MAP ((maximum aposteriori) estimate of $\Theta$. The posterior over $\Theta$ (which needs to be maximized) is given by:

$$\begin{aligned} p(\Theta|R_u) &= p(\Theta)p(R_u|\Theta) \\ &= p(\Theta) \prod_{u \in U} \prod_{i \in B^u} \prod_{j \notin B_u} \sigma(x_{ui} - x_{uj}) \end{aligned}$$

We need to maximize the above posterior function, (or its log-posterior), shown below.

$$\log p(\Theta|R_u) = \sum_u \sum_{i \in B_u} \sum_{j \notin B_u} \ln \sigma(x_{ui} - x_{uj}) - \lambda ||\Theta||^2$$

The first summation term corresponds to the log-likelihood, i.e., $\log p(R_u|\Theta)$ whereas the second term corresponds to the log of the Gaussian-prior, i.e., $\log p(\Theta)$. Here, $\lambda$ is a constant, proportional to $\frac{1}{\sigma^2}$. $||\Theta||^2$ is given by the following expression:

$$||\Theta||^2 = \sum_u ||v_u^{\mathrm{U}}||^2 + \sum_i ||v_i^{\mathrm{I}}||^2$$

The second term is commonly called as the *regularization* term, and is used to prevent overfitting by keeping the learned factors $v_u$ and $v_i$ sparse.

## 2.2 Stochastic Gradient Descent (SGD)

SGD is typically used to optimize objective functions that can be written as sums of (differentiable) functions, e.g., in the objective function above, we have one function per training data point $(u, i, j)$. The standard gradient descent method is an iterative algorithm: Suppose that we want to maximize a given objective function. In each iteration, we compute the gradient of the function and update the arguments in the direction of the gradient. When the function is in the summation form, computing the overall gradient requires computing the derivative for each function in the summation. Hence, computing the derivative can be quite expensive if we have a large training dataset. In SGD, we approximate the derivative by computing it only at a single (randomly chosen) term in the summation and update the arguments in this direction. Despite this approximation, SGD has been shown to work very well in practice [11, 9, 10], often outperforming other methods including the standard gradient descent. We illustrate SGD using the example of the latent factor model. In the above example, a given training data point $(u, i, j)$ defines a term in the summation. The derivatives with respect to the $v_u$, $v_i$ and $v_j$ variables are shown below. Denote $c_{u,i,j}$ to be equal to $(1 - \sigma(x_{ui} - x_{uj}))$.

$$\begin{aligned} \frac{\partial L(U,V)}{\partial \mathbf{v_u}} &= c_{u,i,j}(\mathbf{v_i} - \mathbf{v_j}) - \lambda \mathbf{v_u} \\ \frac{\partial L(U,V)}{\partial \mathbf{v_i}} &= c_{u,i,j}\mathbf{v_u} - \lambda \mathbf{v_i} \\ \frac{\partial L(U,V)}{\partial \mathbf{v_j}} &= -c_{u,i,j}\mathbf{v_u} - \lambda \mathbf{v_j} \end{aligned}$$

Now, we use the above derivatives to update the appropriate factors. Note that since we have a maximization problem, we need to move in the same direction as the derivative. The corresponding update equations are shown below:

$$\begin{aligned} \mathbf{v_u} &= \mathbf{v_u}(1 - \epsilon\lambda) - \epsilon c_{u,i,j}(\mathbf{v_i} - \mathbf{v_j}) \\ \mathbf{v_i} &= \mathbf{v_j}(1 - \epsilon\lambda) - \epsilon c_{u,i,j}\mathbf{v_u} \\ \mathbf{v_j} &= \mathbf{v_j}(1 - \epsilon\lambda) + \epsilon c_{u,i,j}\mathbf{v_u} \end{aligned}$$

Here, $\epsilon$ is the learning rate which is set to a small value. The regularization term $\lambda$ is usually chosen via *cross-validation*. An exhaustive search is performed over the choices of $\lambda$ and the best model is picked accordingly. The overall algorithm proceeds as follows. A training data point $(u, i, j)$ is sampled uniformly at random. The gradients are computed at this particular data point and the variables are modified according to the update rules shown below. An *epoch* is roughly defined as a complete pass over the data set (i.e., over all the non-zero entries in the rating matrix).

## 3. TAXONOMY-AWARE LATENT FACTOR MODEL (TF)

The techniques presented in Section 2 allow for learning a user's personalized ranking over items. However, such a basic model suffers from a number of shortcomings as follows:

- Sparsity: In large scale settings and as often is the case in web applications, the number of items are in the millions and it is often the case that each user rates only a few of them. This creates a sparsity problems and prevents the model from learning that for instance, buying an iPhone is a good indicator that the user would be interested in other electronic gadgets.
- Temporal awareness: User purchases are mainly driven by two factors: short-term interests and long-term interests. Standard latent factor models can capture long-term interests, however they fail to tease apart transient interests from long-term ones.

In this section, we describe our proposed taxonomy-aware temporal latent factor model (TF) for recommendation and illustrate its advantages over previously proposed approaches. Our model resolves the aforementioned shortcomings of the standard latent factor models by incorporating a taxonomy prior over the item factors and modeling time explicitly in the user-item affinity model. We first give some notations and then describe the affinity model. Learning the parameters of the affinity model and recommendation are addressed in Section 4 and 5, respectively.

### 3.1 Notations

We illustrate our TF model using a generative graphical model [15, 23, 12] representation, shown in Figure 2. We define the random variables in the graphical model and then explain the time-specific affinity model in the next subsection. We start off by providing the notations that we use, about the taxonomy.

**Taxonomy notations:**
We use item identifiers using integers $i, j$. Given an item $i$, then $p(i)$ denotes its parent in the taxonomy. $p^m(j)$ denotes the $m^{th}$ node on the path from item $j$ to the root (e.g., $p^0(j) = j, p^1(j) = p(j)$). Also, we use $D$ to denote the depth of the taxonomy.

**Random variables:**
The random variables in the model are shown below. All the random variables except $B_t$ correspond to the factors in the TF model. Each factor is a vector with dimension $1 \times K$, where $K$ is the dimensionality of the model.

- $v_u^U$: denotes the latent factor of the user $u$. ($v_u^U$ is analogous to the user factor $u_i$ for the latent factor model in Section 2). This factor captures the long-term interests of the user.
- $w^I_{p^m(j)}$: We introduce latent variables for all the nodes in the item taxonomy. $w^I_{p^m(j)}$ denotes the latent variable of $p^m(j)$, i.e., the latent variable of the the $m^{th}$ ancestor of item $j$. $w^I_{p^m(j)}$ represents the offset of the node $p^m(j)$ from its parent latent factor, i.e., how much different this category is from its parent.
- $v_j^{\mathrm{I}}$: denotes the effective latent factor corresponding to item $j$. (This factor is analogous to the item factor in the latent factor model in Section 2).
- $v_j^{\mathrm{I}\to\bullet}$: denotes the *next-item* factor, which is used to capture the short-term trends in the user purchases. Intuitively, if a user purchases item $j$, then $v_j^{\mathrm{I}\to\bullet}$ gives a location in the latent space where the next purchase is most likely to occur (this will be made more explicit below). We introduce latent variables $w^{\mathrm{I}\to\bullet}_{p^m(j)}$ for all nodes in the taxonomy in the same manner as above.
- $B_t$: denotes the $t^{th}$ transaction of the $i^{th}$ user, i.e., the set of items that were bought in the $t^{th}$ time instant. The $B_t$ node is shaded in Figure 2 since we know the value of this random variable.

Since we desire the item latent factors to capture the effect of taxonomy, i.e., an item factor should be similar to its ancestors. In particular, we expect the factors of a node to depend on all its ancestors up to the root. We model this as follows:

$$v_j^{\mathrm{I}} = \sum_{m=0}^{D} w^I_{p^m(j)}$$
$$v_j^{\mathrm{I}\to\bullet} = \sum_{m=0}^{D} w^{\mathrm{I}\to\bullet}_{p^m(j)} \qquad (1)$$

Note that this way of defining item factors allows the siblings items in the taxonomy to have similar factors as desired. Also, it allows an item to sufficiently differ from its siblings if that leads to better modeling of the data.

### 3.2 Taxonomy-aware Temporal Affinity Model

For each user and for each time step, we need to determine an affinity array $s$ over all the items $j$ where $s_t(j)$ represents the likelihood that the user $u$ buys item $j$ at time $t$. The score for an item is a sum of two terms.

$$s_t(j) = \langle v_u^U, v_j^{\mathrm{I}} \rangle + \frac{1}{|B^u_{t-1}|} \sum_{\ell \in B^u_{t-1}} \langle v_\ell^{\mathrm{I}\to\bullet}, v_j^{\mathrm{I}} \rangle \qquad (2)$$

1. <u>*Long-term interest:*</u> The first term $\langle v_u^U, v_j^{\mathrm{I}} \rangle$ is the inner product between the user factor $v_u^U$ and the item factor of the $j^{th}$ item, $v_j^{\mathrm{I}}$ which captures the global (long-term) similarity between the user $u$ and the item $j$.
2. <u>*Short-term interest:*</u> The second term in the score is given by $\frac{1}{|B^u_{t-1}|} \sum_{\ell \in B^u_{t-1}} \langle v_\ell^{\mathrm{I}\to\bullet}, v_j^{\mathrm{I}} \rangle$ which captures the short-term similarity, i.e., if the user bought an item $\ell$ (e.g., camera) in the previous time step, then the likelihood that the user buys an item $j$ (e.g., flash memory) in the current time step. In this way, our model captures the temporal dynamics of the user purchase behavior.

Note that both terms are dependent on the user. The first term depends on the user factor while the second term depends on the previous items that were purchased by the user. For each user, we construct the array of scores over all items and this gives us a personalized ranking for what the user will purchase next.

**Higher order affinity models** In the affinity model described in Equation 2, we use the user's transaction at time $t-1$ to predict $B^u_t$. However, using a single time step may not be enough to capture all the temporal dynamics. For instance, user may purchase a camera followed by a flash drive followed by a lens. We also need to capture the dependence between the camera and the lens. In order to do this, we extend the model to use a higher order Markov chain, i.e., we consider $N$ previous transactions to predict the current transaction. We use the following equation to compute the

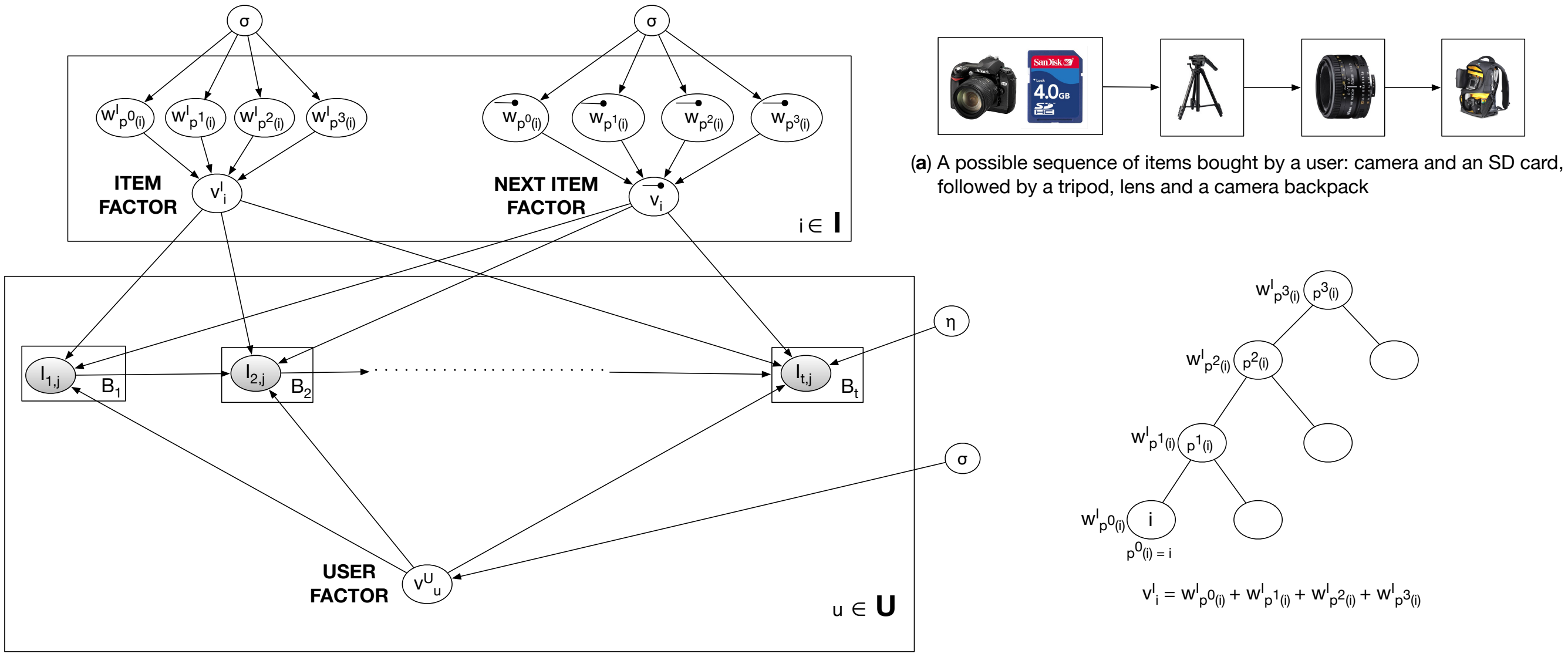

(**a**) A possible sequence of items bought by a user: camera and an SD card, followed by a tripod, lens and a camera backpack

(**b**) TF: Our proposed taxonomy-aware latent factor model

(**c**) Additional latent factors for taxonomy

Figure 2: (a) An example of a sequence of items purchased by the user. Here, the user buys a camera followed by a set of camera accessories. (b) Our proposed taxonomy-aware latent factor model (TF). The user factor, item factor and next-item factor are as shown. The shaded random variables $B_t$ are the items purchased by the user at various times. (c) Illustrating latent factors for the taxonomy. For each node in the taxonomy, we introduce a latent variable, i.e., we introduce the latent variable $w^{\mathrm{I}}_{p^m(i)}$ for the $m^{th}$ ancestor of $i$. The actual item factor is the sum of these variables, as shown in the figure.

scores:

$$s_t^N(j) = \langle v_u^U, v_j^{\mathrm{I}} \rangle + \sum_{n=1}^{N} \frac{\alpha_n}{|B_{t-n}|} \sum_{\ell \in B_{t-n}^u} \langle v_\ell^{\mathrm{I}\rightarrow\bullet}, v_j^{\mathrm{I}} \rangle \quad (3)$$

As shown in the above equation, the first term remains unchanged. In the second term, we extend the summation to $N$ previous transactions of the user. The additional parameters $\alpha_n$ control the extent to which the previous transactions contribute to the current transaction. We use an exponential decay function $\alpha_n = \alpha e^{(-n/N)}$ for this purpose.

# 4. LEARNING

In Section 3 we presented our time-based user-item affinity model. Now, our goal is to learn the parameters of this model to fit the observed user transactions. In other words, at each time step, and for each user, the affinity model should give higher scores to the items the user bought at time $t$ than the affinity scores given to item not bought by the user at time $t$. To achieve this goal we employ the same discriminative training objective function we introduced in Section 2. In Subsection 4.1 we first detail the learning algorithm over the TF model, and then in Subsection 4.2 we give a novel training scheme to further leverage the power of the taxonomy.

## 4.1 Discriminative Training

Now, we describe the algorithm for learning the TF model. As described in Section 3, the parameters we need to learn are the user factors $v_u^U$, item factors $v_i^I$ and the next-item factors $v_i^{\mathrm{I}\rightarrow\bullet}$. To learn $v_i^{\mathrm{I}}$ and $v_i^{\mathrm{I}\rightarrow\bullet}$, we need to learn the latent factors of the taxonomy, i.e., $w_i^{\mathrm{I}}$ and $w_i^{\mathrm{I}\rightarrow\bullet}$ respectively. We use cross-validation to fix the other parameters such as $K$: the dimensionality of the factors, $\sigma$ and $N$ (number of previous transactions to consider) and $\alpha_n$ (decay rate). Denote by $\Theta$, the set of parameters $\{v^U, w^I, w^{\mathrm{I}\rightarrow\bullet}\}$. We exploit recent advances in learning factorization models and rely on the so-called discriminative Bayesian personalized ranking (BPR) which has been shown to significantly outperform generative training [24].

In BPR, the goal is to discriminate between items the user bought and items s/he didn't buy. In other words, we need to learn a ranking function $R_u$ for each user over items such that if item $i$ appears in user u's transaction at time $t$ and item $j$ does not appear in the transaction, then $R_u(i) > R_u(j)$. For this, we need to have: $s_t^N(i) > s_t^N(j)$. Based on the above, our likelihood function $p(R_u|\Theta)$ is given by:

$$p(R_u|\Theta) = \prod_{u \in U} \prod_{B_t \in B^u} \prod_{i \in B_t} \prod_{j \notin B_t} p(s_t^N(i) > s_t^N(j)|\Theta)$$

Here, $B^u$ is all the transactions of the user $u$. As specified in Section 3, we use a Gaussian prior $N(0, \sigma)$ over the entries in $\Theta$. We compute the MAP ((maximum aposteriori) estimate of $\Theta$. The posterior over $\Theta$ (which needs to be maximized) is given by:

$$\begin{aligned} p(\Theta|R_u) &= p(\Theta)p(R_u|\Theta) \quad (4) \\ &= p(\Theta) \prod_{u \in U} \prod_{B_t \in B^u} \prod_{i \in B_t} \prod_{j \notin B} p(s_t^N(i) > s_t^N(j)|\Theta) \end{aligned}$$

We need to maximize the above posterior (or its log-posterior), shown below. Following [24], we approximate the non-smooth, non-differentiable expression $s_t^N(i) > s_t^N(j)$ using the logistic sigmoid function $\sigma(s_t^N(i) - s_t^N(j))$, where $\sigma(z) = \frac{1}{1+e^{-z}}$. Plugging this into Equation (4) and taking the logarithm, we get the following equation.

$$\sum_{u} \sum_{B_t \in B^u} \sum_{i \in B_t} \sum_{j \notin B_t} \ln \sigma(s_t^N(i) - s_t^N(j)) - \lambda ||\Theta||^2 \quad (5)$$

where, $\lambda$ is a constant, proportional to $\frac{1}{\sigma^2}$ and $||\Theta||^2$ is given by the following expression:

$$||\Theta||^2 = \sum_{u} ||v_u^{\mathrm{U}}||^2 + \sum_{i} \sum_{m} ||v^{\mathrm{I}}_{p^m(i)}||^2 + ||v^{\mathrm{I}\rightarrow\bullet}_{p^m(i)}||^2$$

We use stochastic gradient decent to optimize the above function. As described in Section 2, we need to select a sample, i.e., a single term in the summation in Equation 5. For this, we sample a user $u$ and a transaction $B_t^u$. We pick an item $i$ in the transaction and an item $j$ not in the transaction. The four tuple $(u, t, i, j)$ defines a single term in the summation, given by:

$$L(u,t,i,j) = \ln \sigma(s_t^N(i) - s_t^N(j)) - \lambda||\Theta||^2$$
$$= \ln \sigma\Big[\langle v_u^U, v_i^{\mathrm{I}} - v_j^{\mathrm{I}}\rangle + \sum_{n=1}^{N} \frac{\alpha_n}{|B_{t-n}|} \sum_{\ell \in B_{t-n}^u} \langle v_\ell^{\mathrm{I}\to\bullet}, v_i^{\mathrm{I}} - v_j^{\mathrm{I}}\rangle\Big] - \lambda||\Theta||^2$$

Now, we need to compute the gradients of $L(u, t, i, j)$ with respect to the parameters $\Theta = \{v^U, w^{\mathrm{I}}, w^{\mathrm{I}\to\bullet}\}$. Using the chain rule, we note:

$$\frac{\partial L(u,t,i,j)}{\partial w_{p^m(i)}^{\mathrm{I}}} = \frac{\partial L(u,t,i,j)}{\partial v_i^{\mathrm{I}}} \frac{\partial v_i^{\mathrm{I}}}{\partial w_{p^m(i)}^{\mathrm{I}}} = \frac{\partial L(u,t,i,j)}{\partial v_i^{\mathrm{I}}}$$

The second equality follows from Equation 1 ($\frac{\partial v_i^{\mathrm{I}}}{\partial w_{p^m(i)}^{\mathrm{I}}} = 1$). Hence, we can focus on computing gradients with respect to $v^{\mathrm{I}}$ and $v^{\mathrm{I}\to\bullet}$. The gradients of $L(u, t, i, j)$ with respect to these parameters are shown below. Lets us denote $c_{u,i,j,t} = \big(1 - \sigma(s_t^N(i) - s_t^N(j))\big)$, and using the well-known fact that $\frac{\partial \ln \sigma(z)}{\partial z} = (1 - \sigma(z))$, the gradients are given by the following equations. For simplicity, we denote $L(u, t, i, j)$ using $L$.

$$\frac{\partial L}{\partial v_u^{\mathrm{U}}} = c_{u,i,j,t}(v_i^{\mathrm{I}} - v_j^{\mathrm{I}}) - \lambda v_u^{\mathrm{U}}$$
$$\frac{\partial L}{\partial v_i^{\mathrm{I}}} = c_{u,i,j,t}\Big[v_u^{\mathrm{U}} - \sum_{n=1}^{N} \frac{\alpha_n}{|B_{u,t-n}|} \sum_{\ell \in B_{u,t-n}} v_\ell^{\mathrm{I}\to\bullet} - \lambda v_i^{\mathrm{I}}\Big]$$
$$\frac{\partial L}{\partial v_j^{\mathrm{I}}} = -\frac{\partial L(u,t,i,j)}{\partial v_i^{\mathrm{I}}}$$
$$\frac{\partial L}{\partial v_\ell^{\mathrm{I}\to\bullet}} = c_{u,i,j,t}(v_i^{\mathrm{I}} - v_j^{\mathrm{I}}) \sum_{n:\ell \in B_{u,t-n}} \frac{\alpha_n}{|B_{u,t-n}|} - \lambda v_l^{\mathrm{I}} \quad (6)$$

Next, we list the update rules corresponding to the above gradients. For each user factor, we increment it using the user derivative with a small learning rate of $\epsilon$. For the item factors, we need to do $D$ updates, i.e., one for each level in the taxonomy. Note that the gradient with respect to all internal taxonomy item factors is the same due to Equation 1.

$$v_u^{\mathrm{U}} = v_u^{\mathrm{U}} + \epsilon\Big(\frac{\partial L}{\partial v_u^{\mathrm{U}}}\Big)$$
$$w_{p^m(i)}^{\mathrm{I}} = w_{p^m(i)}^{\mathrm{I}} + \epsilon\Big(\frac{\partial L}{\partial v_i^{\mathrm{I}}}\Big)$$
$$w_{p^m(j)}^{\mathrm{I}} = w_{p^m(j)}^{\mathrm{I}} + \epsilon\Big(\frac{\partial L}{\partial v_j^{\mathrm{I}}}\Big)$$
$$w_{p^m(\ell)}^{\mathrm{I}\to\bullet} = w_{p^m(\ell)}^{\mathrm{I}\to\bullet} + \epsilon\Big(\frac{\partial L}{\partial v_\ell^{\mathrm{I}\to\bullet}}\Big) \quad (7)$$

The basic learning algorithm is as follows. We use stochastic gradient descent to train the model. We select a sample which is represented by the 4-tuple $(u, t, i, j)$. Subsequently, we compute the gradients as shown in Equation 6 and subsequently update the factors using the update rules shown in Equation 7. In our experimental analysis, we study the effects of progressively updating the taxonomy, level by level. As described in Section 6, we introduce a parameter `taxonomyUpdateLevels` to update only selected levels of the taxonomy.

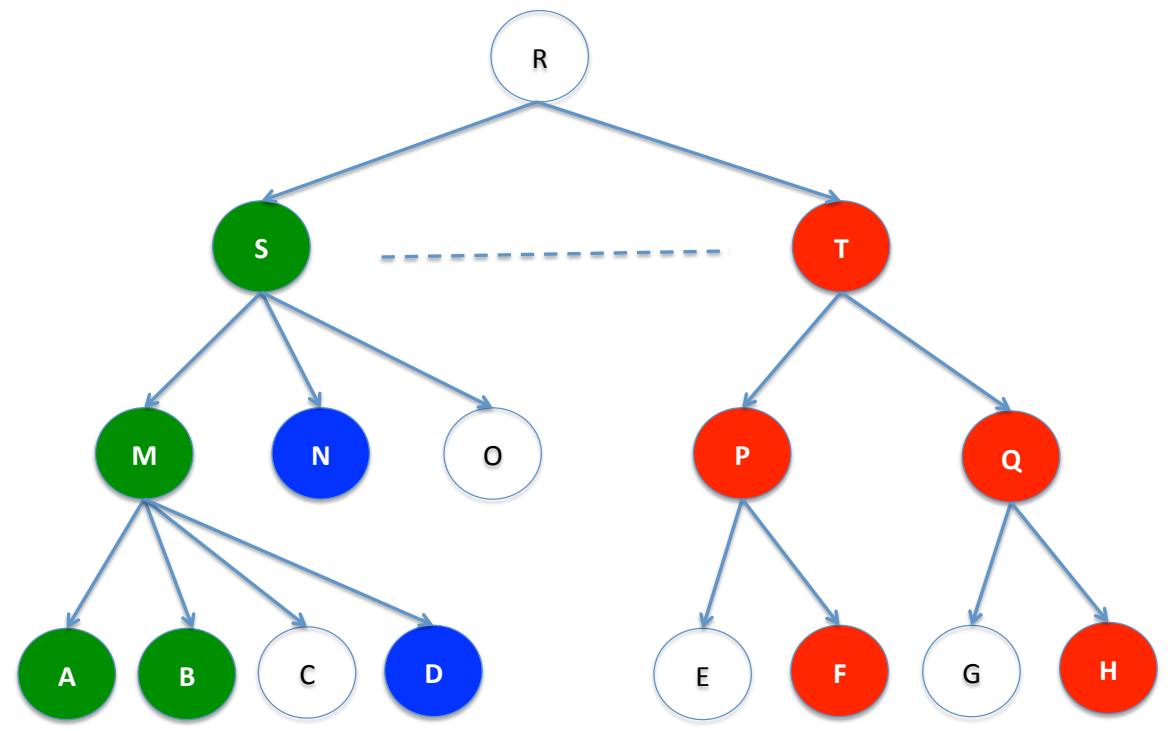

Figure 3: Illustrating sibling-based training. Items in green are bought by the users. Items in red are what the random-sampling algorithm might select. Items in Blue are selected by the sibling-based training algorithm. See text for more details (best viewed in color)

## 4.2 Sibling-based Training

To illustrate the idea of sibling training, we refer the reader to Figure 3. In this figure, we show a taxonomy of depth $D = 3$. We label items (i.e., the leaves of the tree or level 3 nodes) with symbols $A, B, C, \cdots$, the subcategories (i.e., nodes at level 2) with $M, N, \cdots$, and the categories (i.e., nodes at level 1) with $S, T, \cdots$, and finally the root with $R$. As we mentioned in Section 3, each of these nodes are given a factor that we call $w$ with a subscript indicating the node. For example, the factor assigned to node S and A are denoted with $w_S$ and $w_A$ respectively. To compute the effective latent factor associated with a given node, we sum all the factors along the path from this node to the root, as previously described. For example, the effective latent factor of node $A$, $v_A = w_R + w_S + w_M + w_A$, and the effective latent factor of node $N$, $v_N = w_R + w_S + w_N$.

Now suppose that the user bought item $A$. In order to do the training described in Section 4.1, we need to pick another item that the user did not buy and enforce that the user's affinity to item $A$ is larger than the user affinity to the other item. Let us suppose we randomly pick item $F$. Applying the gradient rule over this pair will involve all factors, $W$s, along the paths from the root to each of $A$ (marked with green) and $F$ (marked with red). Intuitively, the gradient update rules should enforce the fact that the given user prefers category $S$ to category $T$, subcategory $M$ to $P$, and $A$ to $F$. Now consider that the user bought item $B$, therefore, we need to sample another item not bought by the user. Let us say we randomly sample $H$. Using the same logic as above, the gradient update rules should affect all factors $W$s, along the paths from the root to both $B$ and $H$. The outcome of these updates would be to enforce that the user prefers category $S$ to category $T$, subcategory $M$ to $Q$, and $B$ to $H$. However, there is a redundancy involved here. We already know from the first update that the user is more interested in the subtree rooted at $S$ than the subtree rooted at $T$, moreover, while we know that the user is interested in the subtree rooted at $S$, we did not model the fact that user prefers the subtree rooted at $M$ to the subtree rooted at $N$.

To remedy this deficiency, we introduce a novel training criteria that we call sibling-based training. The idea here is that while random sampling enforces users preference coarsely over the taxonomy, it fails to fine-tune those preferences over the sibling nodes in the taxonomy. Thus, we proceed as follows to solve this problem. Suppose, as before, that the user bought item $A$, then we need to enforce that the user prefers each node along the path form $A$ to the root over its siblings. That is, the user prefers $A$ to $B, C, D$; $M$ to $N, O$; and $S$ to $T$. To do that, we apply the BPR gradient update rule to each internal node along that path. First we pick a sibling node of $A$, say $D$ (marked with blue in Figure 3), and enforce that the user affinity to $A$ is larger than his/her affinity to $D$. Similarly, we pick a sibling node of $M$, say $N$ (marked with blue in Figure 3), and enforce that the user affinity to $M$ is larger than the user's affinity to $M$. And finally repeat the process at level 1. At each of these steps, the gradient update rule will affect all the factors $W$s from the level under consideration up until the root. Thus, each observed purchase from the user will result in $D$ training examples. In our implementation, we mix random sampling with sibling-based training to reap the benefits of each of them: coarse learning from the random sampling and fine-tuning ability of the sibling-based training. As we will demonstrate in the experimental section, this sibling-based approach results in significant improvement over random-based training.

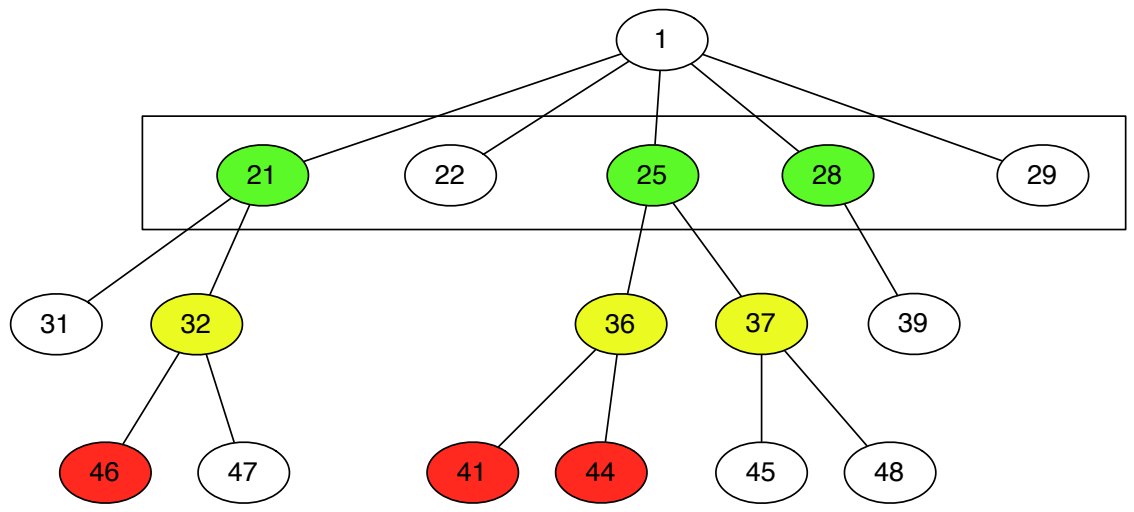

Figure 4: The figure pictorially illustrates our cascaded inference procedure. In the first step, we rank the top-level of the taxonomy, shown in the rectangular box. We pick the top-3 items (here, they are items 21, 25 and 28). Next, we consider the children of these items ONLY and continue the recursion. Finally, we return the top items (shown in red) at the leaf level.

# 5. RECOMMENDATIONS

In this section, we describe how to exploit the taxonomy for speeding up inference in latent-factor models. As specified in Section 1, we are interested to compute the top-$k$ products for each user. As with traditional latent factor models, we need to first compute the affinity of a user to all the items in the dataset and subsequently select the best $k$ among them to recommend to the user. As we describe in Section 7.1, we have about 1.5 million products in our shopping dataset and computing the top-$k$ items for each user requires millions of vector operations (Equation 3) per user. Hence, this approach does not scale to large number of users and items. Next, we illustrate our proposed cascaded inference algorithm, that exploits the taxonomy to speedup the inference.

## 5.1 Cascaded Inference

The overall idea is depicted in Figure 4. In this figure we depict the taxonomy where leaf nodes are product and internal nodes are categories (and sub-categories). As we noted earlier, each node in the tree is associated with a factor $w^{\mathrm{I}}$ which models its offset in the latent space from its parent. The products' factors are then computed by the additive model in Equation 1. Taking this idea a step further, we can also construct an effective factor for each internal node (category) by summing all offsets in the path from that node to the root. Therefore we can perform recommendation at any level in the tree by ranking that level's factors according to their affinities to the user factor under consideration.

To perform recommendation at the leaf level we have to compute the ranking over millions of items, thus to speed this procedure we use the taxonomy to prune the search space as follow. We noticed (see Figure 7(e)), that the taxonomy imposed a clustering effect on the latent factors and that the magnitude of the values of the offset of nodes form their parents decreases as we move down the tree. Therefore, we employ the following top-down algorithm. We start the algorithm by considering the nodes in the top most level in the taxonomy in our search space. We rank them according to our affinity criterion (Equation 3). Next, we pick the top-$k_1$ items in this level and include the children of these items in our search space. We compute the scores of these items, select the top-$k_2$ items, and continue the recursion along their children. The benefits of this approach are two-fold: first, our search space is guaranteed to be small, resulting in much better efficiency, second we provide a more semantically meaningful ranking over the items as we can stop the recursion at any level and provide recommendation at that level if desired. Cascaded inference naturally provides us with a trade-off between accuracy and efficiency. As we increase the values of $k_i$'s, we improve accuracy at the expense of additional computation. We experimentally demonstrate this trade-off in Section 7.5.

# 6. IMPLEMENTATION DETAILS

We developed a multi-core implementation of our taxonomy-aware latent factor model in C++. We used the BOOST [1] library package for storing the factor matrices and for locks. Our implementation of TF is generic, i.e., we can simulate a wide variety of previously proposed models include latent factor models and FPMC [25], which is the current state-of-the-art in next-transaction recommendation. We parameterize our implementation using the following parameters:

- `taxonomyUpdateLevels`: This is a parameter that controls the number of levels of the taxonomy that we use in our training/inference (going up from the leaf level of the taxonomy). For instance, if `taxonomyUpdateLevels = 1`, then we only use the product level, just as in traditional latent factor models. Similarly, if `taxonomyUpdateLevels` = 4, then we use the full taxonomy. The benefit of this parameter is to test the effect of adding more depth to the taxonomy on the quality of the recommendations.
- `maxPrevtransactions`: This parameter represents the number of previous transactions to use when predicting the user's short-term interest, i.e., the order of the Markov chain.

Note that if we set `taxonomyUpdateLevels` to 1 and the `maxPrevtransactions` to 0, then we only update the last (leaf) level in the taxonomy and ignore time – which is equivalent to the latent factor model. On the other hand, the configuration where `taxonomyUpdateLevels` = 1 and `maxPrevtransactions` = 1 corresponds to the FPMC technique. In order to scale our learning algorithms to very large amounts of users/items, we develop techniques to parallelize our training and testing algorithms. We start with the a description of the parallelization in train.

## 6.1 Parallelizing Training

We develop a multi-threaded approach using locks. The global state maintained by the SGD algorithm are the 3 factor matrices $\{v^U, v^I, v^{\mathrm{I}\rightarrow\bullet}\}$. We introduce a lock for each row in our factor matrices. As shown in Section 4, in each iteration of training, we execute 3 steps: in the *first step*, we sample a 4-tuple $(u, i, j, t)$; in the *second step*, we compute the gradients with respect to the appropriate user and item factors. In the *third step*, we update the factor matrices based on the gradients.

In the second step, we read the item factors. Hence, we need to obtain a read-lock over the factor and release it after reading. In the third step, we write to the factor thus we need to obtain a write lock on the item factor and subsequently release the lock once we update the factor. As shown in Section 4, for each iteration of SGD, we need to update the item factors of all of its ancestors and possibly all the item factors of its previous transactions. As we describe in Section 7, the taxonomy used in the evaluation has 1.5 million items at the last level and about 1500 internal nodes. Hence, these nodes are updated much more frequently (around 1000 times more often) than the items at the last level. Therefore we expect much more contention for the factors corresponding to the internal nodes in the taxonomy. To handle this issue, we propose the following caching technique. Here, each thread maintains a local cache of the item factors which correspond to the internal nodes in the taxonomy. Now, in the second step of each SGD iteration, the factors are read from the global copy and in the third step, the local cached copy is updated. Whenever the difference between the corresponding local and global copies exceeds a threshold, we reconcile the local cached copy with the global factor matrices.

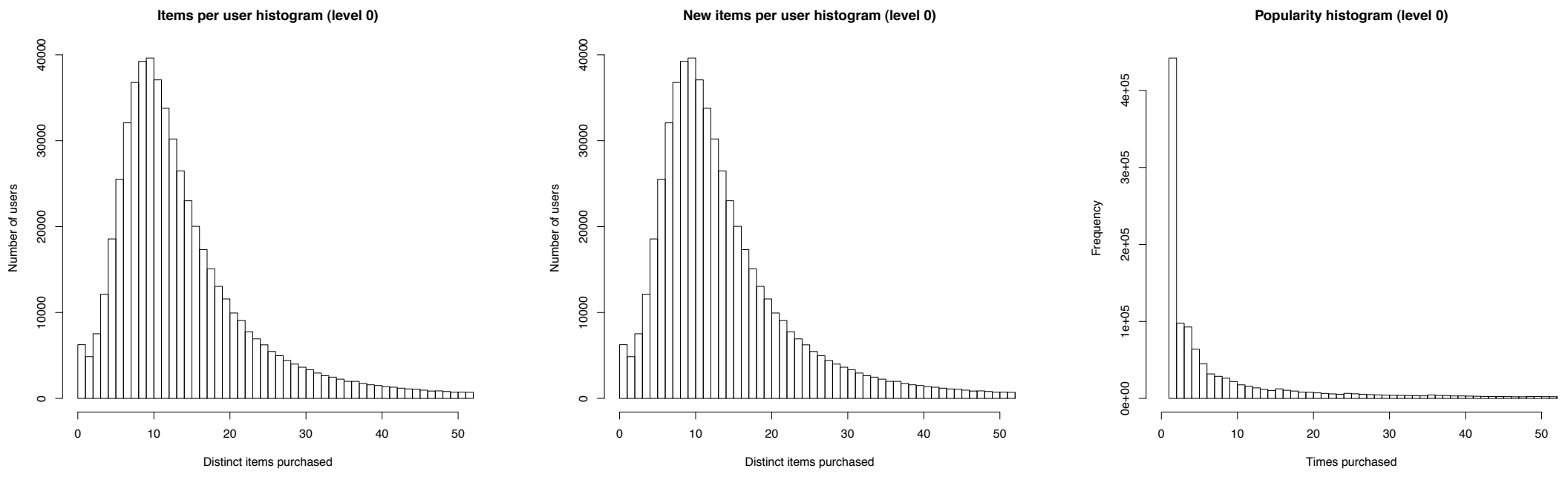

(a) # Distinct items (b) # New items (c) Item popularity

Figure 5: The figures illustrate the characteristics of our data set. Part(a) shows the number of distinct items bought by the user (train). Part(b) illustrates the number of new items that the user (test), while part (c) illustrates the item popularity.

## 6.2 Parallelizing Evaluation

Next, we propose techniques to parallelize our evaluation and computing the metrics. This is particularly significant because we need to do several rounds of cross-validation to compute the values for the parameters $(\lambda, K)$ of the stochastic gradient descent. Essentially, we partitioned the set of users into several machines using Hadoop and on each machine, we ran the evaluation and aggregated the results.

# 7. EXPERIMENTAL EVALUATION

In this section, we illustrate the benefits of using TF models over the state-of-the-art approaches in recommender systems – in terms of accuracy and efficiency. We start with a description of our dataset, metrics and related systems against which we compare our proposed TF model.

## 7.1 Dataset

To evaluate the TF model, we used a log of user online transactions obtained from a major search engine, email provider and online shopping site. The dataset contains information about the historical purchases of users over a period of 6 months. We fully *anonymize* the users by dropping the original user identifier and assigning a new, sequential numbering of the records. We drop the actual time stamp and only maintain the sequence of the transactional purchase logs. For reasons involving non-disclosure, we report results over a sample of the above data. In the sample, we have about 1 million anonymized users with an average of 2.3 purchases per user and 1.5 million distinct individual products, which is mapped to the Yahoo! shopping taxonomy. The (resulting) taxonomy has about 1.5 million individual products in the leaf level organized into a taxonomy 3 levels deep, with around 1500 nodes at lowest level, 270 at the middle level and 23 top level categories.

For each user, we pick a random fraction of transactions (with mean $\mu$ and variance $\sigma$) and select all subsequent (in time) transactions into the test dataset. All the previous transactions are used for training. To simulate sparsity, we experiment with different values of $\mu = \{0.25(\text{sparse}), 0.50, 0.75(\text{dense})\}$. We use a small variance of $0.05$. Unless otherwise specified, we use $\mu = 0.5$ in all experiments. The last $T$ transactions in the training dataset are used for cross-validation and first $T$ transactions in the test dataset is used for prediction and reporting the error estimates. In all experiments, we use $T = 1$. Since our goal is to build a recommender system which is supposed to help users discover new items, we remove those items (*repeated purchases*) from the users' test transactions which were previously bought by the user, i.e, they appear in the users' train transactions (which most recommender systems can easily do anyway).

In order to better analyze the results of our evaluation, we study the characteristics of the data, pertaining to sparsity. We measure statistics of the number of items purchased by the users and the popularity of the items. In the histogram shown in Figure 5(a), we depict the number of users who bought $k$ distinct items (in the training data) for different values of $k$. The dataset is quite sparse since a very small fraction of the users buy a large number of items. In Figure 5(b), we show the the number of new items purchased by the users, in the test data. This shows that users bought several new items in the test dataset. In Figure 5(c), we depict the item popularity, i.e., we show the number of distinct users who purchased a given item. Notice the heavy tail in the items purchased.

## 7.2 Systems

- **$\mathbf{MF(B)}$:** Latent factor model where B is the previously introduced `maxPrevtransactions` parameter. We compare our TF model against the basic latent factor model. For a fair comparison, we use the BPR algorithm to train the MF model as well, using the setup described in Section 2. In addition, we vary the number of `maxPrevtransactions` parameter from $\mathbf{B} = \{0, 1, 2, 3\}$. Note that $\mathbf{MF(0)}$ corresponds to the SVD++ model of Koren et al. [19] and $\mathbf{MF(1)}$ corresponds to the FPMC (factorized personalized Markov chains) model of Rendle et al. [25], which is the current state-of-the-art technique.

- **$\mathbf{TF(U, B)}$:** This is our proposed taxonomy-aware latent factor model. It is parameterized using two parameters: $\mathbf{U}$ de-

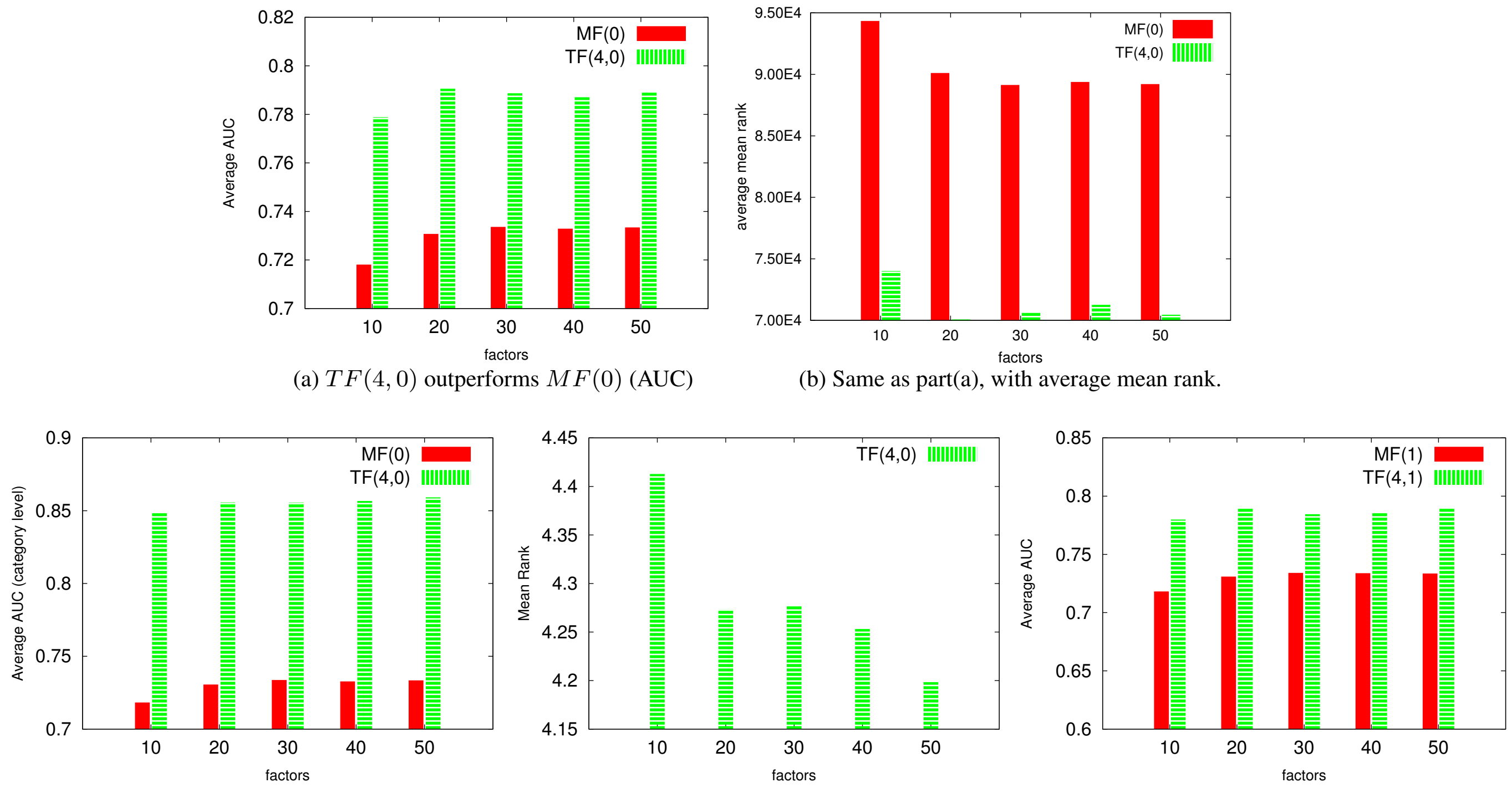

(a) $TF(4,0)$ outperforms $MF(0)$ (AUC)

(b) Same as part(a), with average mean rank.

(c) $TF(4,0)$: AUC at category level

(d) $TF(4,0)$: mean rank at category level

(e) $TF(4,1)$ outperforms $MF(1)$ (AUC)

Figure 6: In this figure, we illustrate the benefits of using the TF model over the baseline models. In part(a,b) we illustrate the benefit of TF(4,0) over the standard MF(0) model. In part(a), we present AUC values and show that improvements of more than 6% are achieved using the taxonomy. In part(b) we illustrate the average mean rank. TF models allow recommendation at the category level. In part(c), we show the AUC of the TF(4,0) model at the category level, in comparison with the MF model. As shown in part(d), the average mean rank of the TF(4,0) model at the category level is around 4. In part(e), we show that using $TF(4,1)$ provides higher `AUC` than $MF(1)$.

notes the `taxonomyUpdateLevels` parameter (i.e. the number of super-categories used from the taxonomy, for example for $U = 1$, no part of the taxonomy is used, just like MF and for $U = 2$ only the sub-categories above the item level are used) and $B$ denotes the `maxPrevtransactions` parameter (Section 6). In our experiments, we use **U** $=\{$`2, 3, 4`$\}$ and **B** $= \{0, 1, 2, 3\}$.

## 7.3 Metrics

We use the metrics described below to compare our model with the above systems.

- <u>`AUC`:</u> (Area under the ROC curve)
`AUC` is a widely used metric for testing the quality of rank orderings. Suppose the list of items to rank is $X$ and our test transaction is $T$. Also suppose $r(x)$ is the numerical rank of the item $x$ according to our model (from $1 \dots n$). Then, the formula to compute `AUC` is given by: (Here $\delta(\phi)$ is the indicator function that returns 1 if $\phi$ is true or 0 otherwise)

$$\frac{1}{|T||X \setminus T|} \sum_{x \in T, y \in X \setminus T} \delta(r(x) < r(y))$$

- **Average** `meanRank`
While the `AUC` metric is widely adopted in practice, it is insensitive to the ranking in our case since we have over 1 million products in our dataset. Suppose we have exactly 1 item in the test transaction and its rank (e.g., according to MF) is $10,000$. The `AUC` value is approximately 0.99. However, if the item's rank is 100 (which is clearly much better than 1000), the `AUC` value is 0.999. For better comparison between the different approaches, we also measure the average `meanRank`. We compute the average rank for each user and them average this value across all users. The average meanRank is related to *f-measure* metric (harmonic mean of precision and recall) metrics, which are commonly used in information retrieval.

Next, we present the results of our experimental analysis. We illustrate the experimental results relating to the accuracy of the modeling in Section 7.4 and the results relating to the efficiency of our models in Section 7.5.

## 7.4 Accuracy Results

In this section, we show results about the benefits of using TF models for improving accuracy of recommendation.

### 7.4.1 *Accuracy Improvement over Baseline Models*

**<u>Improvement over MF(0) model</u>**
In the first experiment, we compare the accuracy of $TF(4,0)$ and $MF(0)$ models. Recall from Section 7.2, that $TF(4,0)$ indicates that we use all the levels of the taxonomy (taxonomyUpdateLevels = 4) and we do not use any previous transactions in the Markov chain (maxPrevtransactions = 0). Since the number of free parameters in $TF(4,0)$ and $MF(0)$ are different, we experiment with a range of factors (to effectively carry out model selection for each model). We investigate the performance of the models with $\{10, 20, 30, 40, 50\}$ factors. We measure the average AUC for the two approaches. The results are shown in Figure 6(a). As shown in Figure 6(a), the best average AUC for $TF(4,0)$ (which occurs at 20 factors) is higher than the best average AUC for $MF(0)$ (also occurring at 50 factors). We would like to note here that both the models above use the same amount of information: they do not take into the account the actual sequence of items purchased, rather they only consider the set of items that has been purchased.

As described in Section 1, using a taxonomy allows us to rank not only at the product level, but also at the category levels. In Figure 6(c), we show the average AUC at the category level for the $TF(4,0)$ model for various factor sizes. For comparison, we also indicate the AUC of the $MF(0)$ model at the product level. As shown in the figure, the taxonomy-aware factor model greatly outperforms the matrix factorization approach.

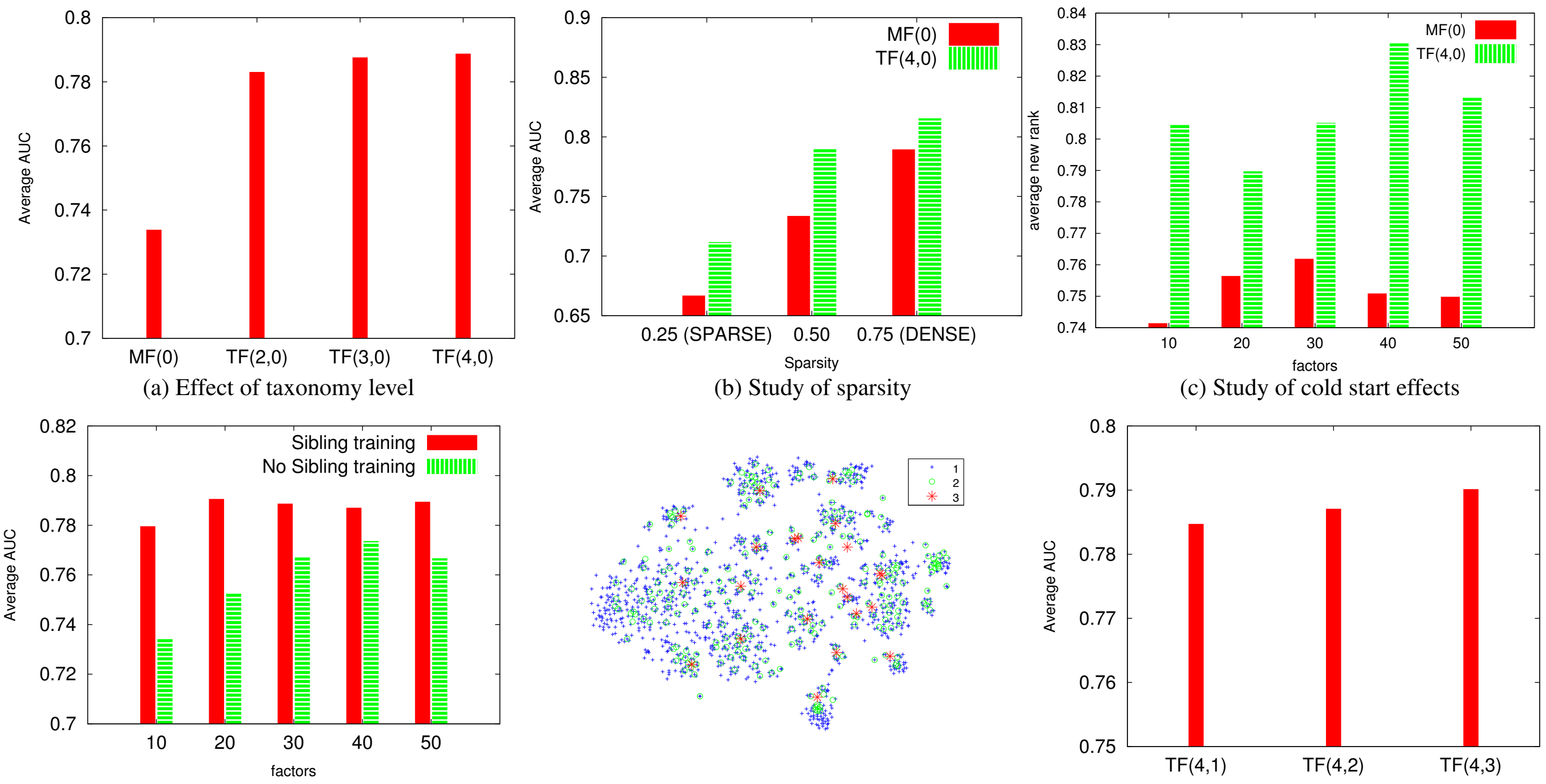

(a) Effect of taxonomy level (b) Study of sparsity (c) Study of cold start effects

(d) Effect of Sibling training (e) Visualizing 2-d projection of factors (f) Effect of time

Figure 7: In this figure, we study the effects of taxonomy. (a) As we increase the number of levels, we improve the average AUC. (b) When the data set is sparse, usage of taxonomy provides much greater benefits. (c) $TF$ models provide better AUC for items with cold start by exploiting taxonomy. (d) Sibling training techniques improve AUC. In part(e), we project the learned factors on to a 2-dimensional plane; notice that the item factors occur close to their ancestors. (f) AUC improves as we increase the order of the Markov chain.

In Figure 6(b), we indicate the meanRank for the $MF(0)$ and the $TF(4,0)$ models at the product level. As shown in the figure, the mean rank for the taxonomy models is an order-of-magnitude less than that for the matrix factorization model. In Figure 6(d), we indicate the mean rank at the category level for the $TF(4,0)$ model. As shown in the figure, the mean rank is just 4, illustrating the benefits of using taxonomy for recommendation. We note here that even though the average mean rank at the product level is high (about 10,000), the average mean rank at the category level is very small and is practical to use.

**Improvement over MF$(1)$ model**

In the second experiment, we compare the accuracy of $TF(4,1)$ and $MF(1)$ models. Recall from Section 7.2 that $MF(1)$ corresponds to the matrix factorization model along with a single step Markov chain (maxPrevtransactions = 1), which also corresponds to the state-of-the-art FPMC technique of Rendle et al. [25]. We compute the average AUC for the two approaches and show the results Figure 6(e). As shown in the figure, using taxonomy significantly improves the recommendation accuracy.

### 7.4.2 *Study of Taxonomy over Recommendation*

**Effect of Taxonomy over recommendation accuracy**

In this experiment, we plan to understand the effect of taxonomy over the recommendation accuracy. We compare the following models: $MF(0), TF(2,0), TF(3,0), TF(4,0)$. We show the results in Figure 7(a). As shown in the figure, the accuracy increases as we incorporate more levels in the taxonomy. This is because of the additional latent factors that we introduce, based on the taxonomy enables better sharing of the statistical strength across items.

**Tackling sparsity and cold start problems**

As discussed in Section 1, the two key problems with state-of-the-art systems in collaborative filtering is sparsity and cold start problems. We simulate sparsity by generating multiple datasets with different values of the split parameter $\mu$. For each dataset, we train the $MF(0)$ and the $TF(4,0)$ models and measure the accuracy of prediction using AUC. We present the results in Figure 7(b). As shown in the figure, for all the three data sets (from sparse to dense), the taxonomy model outperforms the matrix factorization models. Also, note that the benefit of using the taxonomy is very large when the data is sparse. As shown in the figure, for the sparse dataset, we obtain more than a 5% improvement in the AUC for the highly sparse dataset and about 2% for the less sparse dataset.

Next, we show how the taxonomy helps in ameliorating the *cold start* problem. For each of the *new items*, items which do not appear in the training dataset (i.e., they are released at a later time), we measure the average rank of these items, over all the times it was purchased. We compare $MF(0)$ and $TF(4,0)$ for this experiment. As described in Section 3, while the ranking of $MF(0)$ is completely random; in TF-based models, we use the item's immediate super-category as an estimate for its factor. We plot the average `rank` of the new items in Figure 7(c). As shown in the figure, for new items, TF models outperform the $MF(0)$ approach for almost all factor sizes.

**Performance of sibling training**

In this experiment, we evaluate the benefits of using the sibling based training for training the $TF$ models. We train the $TF(4,0)$ models with and without the sibling training and compare the resulting prediction performance. We show the results in Figure 7(d).

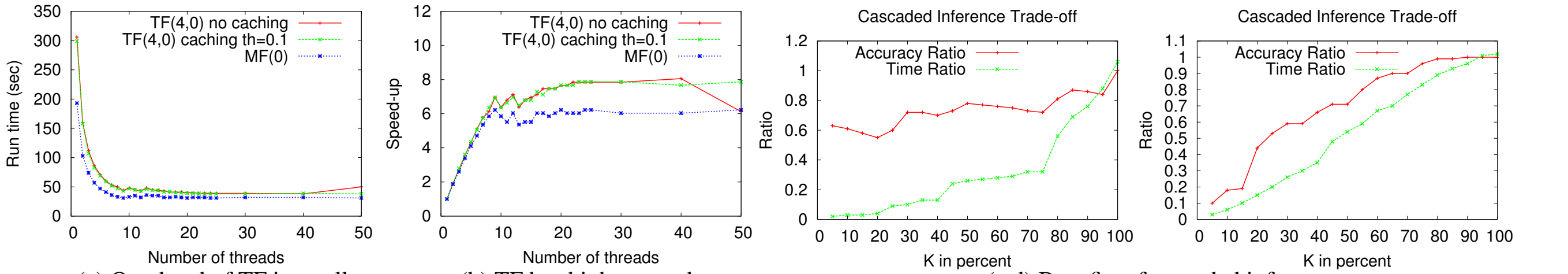

(a) Overhead of TF is small (b) TF has higher speed-up (c,d) Benefits of cascaded inference

Figure 8: (a) Here, we indicate the time per epoch for $TF(4,0)$ and $MF(0)$. The overhead of training the TF model is much less as we increase the number of threads. (b) In this figure, we show the speedups of the above models. Note that $TF(4,0)$ has a maximum speedup of 8 which is higher than $MF(0)$'s maximum speedup of 6. In parts (c,d), we show the trade-offs obtained using our cascaded inference algorithms. As shown in part(c), we can achieve 80% accuracy in 50% of the computation time.

As before, we depict the results for various factor sizes. As shown in the figure, using sibling training improves the overall AUC by more than 3%.

**Visualizing factors in $TF$ model**

Next, we examine the structure of the factors. We use t-SNE [28] a tool that can be used to perform dimensionality reduction on the factors. We plot the item factors (Section 3) on a 2-dimensional plane. In Figure 7(e) we plot the upper 3 levels of the taxonomy. As shown in the figure, the red color points correspond to the topmost level of the taxonomy, similarly the green and the blue color points correspond to levels 2 and 3 of the taxonomy. The factors are clustered into various clusters as expected. Also notice that each red point (corresponding to the topmost level) is surrounded by a set of green points (level 2), which in turn is surrounded by the blue points (level 3), indicating the role that the taxonomy plays in constraining such factors together, i.e., items being close to their children. We also note that some of the factors corresponding to lower levels have moved away from their ancestor factors; we attribute this to the data. Apriori, the factors are similar to their ancestor factors. After an item is purchased a few times and becomes popular, its buying pattern no longer matches that of its parent and therefore its factor gets modified such that it moves away from its ancestors.

### 7.4.3 *Improvement with Higher Order Markov Chains*

In this experiment, we study the benefit of using higher order Markov chains. We compare the accuracy of $TF(4,0)$, $TF(4,1)$ and $TF(4,2)$: which progressively increases the number of previous transactions considered, i.e., the order of the Markov chain (Section 3). We plot the corresponding AUC values in Figure 7(f). As shown in the figures, the prediction error increases as we increase the order of the Markov chain.

## 7.5 Performance Results

In this section, we present the results pertaining to the efficiency of our approaches against naive methods. Specifically, we present the speedups in learning that we are able to obtain using multiple cores and the benefits of using our cascaded inference algorithm.

**Training parallelization**

We investigate the performance of our multi-core implementation of the training algorithm. For the models $\{MF(0), TF(4,0)\}$, we measure the time taken per epoch. For this experiment, we used a 12 core machine running Linux. Note that epoch is a fixed number of iterations for both models. We plot the results in Figure 8(a,b) as a function of the number of threads used. Figure 8(a) shows the absolute wall-clock times and (b) indicates the *speedup*. Clearly, $TF(4,0)$ is more expensive since it has to update the higher levels of the taxonomy as well. We make the following observations. First, as we increase the number of cores, we observe that we obtain an almost linear speed-up initially. Subsequently, the speed-up flattens out as expected. Second, we observe that the maximum speedup for $TF(4,0)$ is 8, whereas the maximum speedup for $MF(0)$ is only 6. Since $TF(4,0)$ also needs to update the higher levels of the taxonomy (of which there are only about 2000 nodes), we expect it to be bottlenecked by the lock contention at this level. However, surprisingly, $TF(4,0)$ has a higher speedup than $MF(0)$. This behavior can be explained if we note that $TF(4,0)$ requires more computation per thread than $MF(0)$ and thus the rate of updating the shared factors per thread is less than the same rate from $MF(0)$. Therefore, to reach the same bottleneck update rate of $MF(0)$, $TF(4,0)$ requires more threads before its speed-up starts to asymptote. As a result of this, we see that when we use a reasonable number of thread (about 10), using $TF(4,0)$ has very little additional overhead over using $MF(0)$. As shown in the figure, as the number of threads is increased, the gap between $MF(0)$ and $TF(4,0)$ reduces to almost zero. Next, we investigate the benefits of our caching techniques in improving speedup. As shown in Figure 8(b), we see that after 40 threads, while the speedup without caching drops, we continue to see a constant speedup with caching. However, we notice that caching only helps when we have more than 40 threads. This is because the bottleneck due to lock contention over the upper levels of the taxonomy occurs only with 40 threads while the bottleneck due to the number of cpu cores occurs earlier, around 10 threads.

**Cascaded Inference**

As illustrated in Section 5.1, cascaded inference provides a trade-off between accuracy and performance. As we increase the search space (increase the values of $n_1$, $n_2$, $n_3$ (Section 5.1)), we improve accuracy at the expense of computational efficiency. In this experiment, we study this trade-off empirically. Since the number of nodes in each level of the taxonomy is different, we use the $k_i$ to denote the percentage of the number of nodes in level $i$ that is considered, i.e., $n_i = k_i size(i)$, where $size(i)$ denotes the number of nodes in level $i$. We evaluate two models of cascading: In the first model, we increase the values of all of $k_1$, $k_2$ and $k_3$ from 0 to 100% gradually. To illustrate the trade-off, we plot two quantities. First, we plot the ratio of the AUC obtained using cascaded inference against the actual AUC (which we computed using the naive method). Second, we also plot the ratio of the computation times between cascaded inference and the naive method. The results are shown in Figure 8(c). As we increase $k$, note that the AUC values increase and the time taken also increases. Even at around 50% of the time consumed, we can achieve close to 80% of the accuracy. In the second model, we hold $k_1$ and $k_2$ to the maximum possible value (the size of the level) and increase $k_3$ from 0 to 100%. We observe a similar trade-off with this model as shown in Figure 8(d). While in the first model, we obtain a non-monotone behavior of the accuracy curve (since we are modifying the number of internal nodes used), we observe a monotonically increasing accuracy in the second approach. Since we only append to the set of items considered, the AUC values increase gradually.

# 8. RELATED WORK

**FPMC model (Next-basket recommendation)**

Rendle et al. [25] proposed the FPMC model which is a combination of the latent factor model and the Markov chain model for predicting the next basket that will be purchased by the users. Our proposed TF model is a more general model and it subsumes FPMC and other models (by setting the `taxonomyUpdateLevels` parameter to 1 and `maxPrevtransactions` to 1, we can recover FPMC). As shown in our experimental analysis, TF outperforms FPMC by exploiting the additional features of the taxonomy.

**Taxonomy-aware latent factor models**

In the recently concluded KDDCup 2011 [2], Yahoo! released a portion of the Yahoo! Music ratings dataset with information about songs, genres, albums and artists which form a taxonomy. Mnih et al. [22] and Menon et al. [21] propose to use the taxonomy to guide learning. The purpose of using the taxonomy was to account for the biases involved in the ratings (since users were allowed to rate songs, genres and artists – multiple levels of the taxonomy) and not particularly to improve recommendation accuracy. In our work, we propose a principled approach to incorporate taxonomy data into the latent factor model framework for improving the accuracy of the prediction and efficiency of the training. In addition, we present the sibling-based training algorithm which enables us to fully exploit the taxonomy. Further, we also present the cascaded inference algorithm for improving the efficiency of inference. Earlier work by Ziegler et al. [30] and Weng et al. [29] also incorporate taxonomies for recommendation using alternative approaches.

**Association rule mining & variants**

Frequent itemset mining and association rule mining [14] techniques have been extensively studied in the database literature over the past couple of decades and led to the foundation of *data mining*. In their seminal work, Agrawal et al. [5] developed the Apriori algorithm for efficiently extracting *sets* of items that are purchased together. In subsequent work [6, 27], the above algorithm has been extended to mine temporal sequences and also incorporate a taxonomy of items into the Apriori algorithm. As shown by its success in the Netflix prize [8], latent factor models typically outperform these techniques for personalized recommendations.

# 9. CONCLUSIONS

User personalization and targeting are key problems in Computational Advertising. Developing accurate recommendation algorithms is therefore key to learning user preferences and providing content that is of interest, to the users. In this paper, we develop a system for product recommendations to the users based on historical purchase logs. However, much of the data is sparse (too few purchases and too many items) and several new items are released each day (leading to cold start issues). Also, most of the user feedback is implicit, i.e., we know of a purchase, but do not know about the users' ratings for the product. In our work, we propose a principled approach for combining latent factor models and taxonomies to resolve the above challenges. Taxonomies are available for large number of products including music (genre, artist, album), movies (genre, director) and we exploit such additional data to develop richer models for recommendation. We develop efficient algorithms to learn taxonomy-based models and also show how to improve the efficiency of our inference algorithms by making use of the taxonomy. Our comprehensive experimental analysis illustrates the benefits of our approaches for product recommendation.